\documentclass[a4paper,12pt]{article}
\usepackage{amssymb,graphicx}
\usepackage{epstopdf}
\usepackage{amsmath,amsfonts,amsthm}
\usepackage{cases}
\usepackage{cite}
\usepackage{epsfig}
\usepackage[symbol*]{footmisc}

\usepackage[pdfencoding=auto]{hyperref}
\usepackage{xcolor}
\usepackage{hyperref}
\definecolor{linkcolor}{HTML}{799B03}
\definecolor{urlcolor}{HTML}{799B03}
\hypersetup{pdfstartview=FitH,linkcolor=linkcolor,urlcolor=urlcolor,colorlinks=true}

\begin{document}

\begin{flushright}
INR-TH-2016-024
\end{flushright}

\vspace{10pt}

\begin{center}
{\LARGE \bf Cosmological bounces and Lorentzian} 

\vspace{0.4cm}

{\LARGE \bf wormholes in Galileon theories}

\vspace{0.4cm}

{\LARGE \bf with an extra scalar field}

\vspace{20pt}

R. Kolevatov$^{a,b}$\footnote[1]{\textbf{e-mail:} kolevatov@ms2.inr.ac.ru} and
S. Mironov$^{a}$\footnote[2]{\textbf{e-mail:} sa.mironov\_1@physics.msu.ru}

\vspace{15pt}

$^a$\textit{Institute for Nuclear Research of the Russian Academy of Sciences,\\
60th October Anniversary Prospect, 7a, 117312 Moscow, Russia}\\

\vspace{5pt}

$^b$\textit{Department of Particle Physics and Cosmology, Physics Faculty,\\
M.V. Lomonosov Moscow State University,\\
Vorobjevy Gory, 119991 Moscow, Russia}
\end{center}

\vspace{5pt}

\begin{abstract}
We study whether it is possible to design a ``classical'' spatially flat bouncing cosmology or a static, spherically symmetric asymptotically flat Lorentzian wormhole in cubic Galileon theories interacting with an extra scalar field. We show that bouncing models are always plagued with gradient instabilities, while there are always ghosts in wormhole backgrounds.
\end{abstract}

\section{Introduction}
Generalized Galileon theories are capable of violating the null energy condition (NEC) without obvious pathologies (ghosts, gradient instabilities, etc.). Galileon Lagrangians are quite peculiar: they involve second derivatives, but the corresponding field equations are nevertheless of the second order \cite{Horndeski:1974wa,Fairlie:1991qe,Fairlie:1992he,Fairlie:1992nb,Nicolis:2008in}.

The original Galileon theories have been generalized in various ways and used for constructing nontrivial solutions within general relativity \cite{Deffayet:2009wt,Chow:2009fm,Deffayet:2009mn,Deffayet:2010zh,Padilla:2010de,Kobayashi:2010cm,Hinterbichler:2010xn,Padilla:2010tj,Goon:2010xh,Deffayet:2011gz,Kobayashi:2011nu,Hinterbichler:2012yn}; for a review, see Ref. \cite{Rubakov:2014jja}. Among the applications of Galileon theories are, for example, the Genesis cosmological scenario \cite{Creminelli:2010ba}, bouncing universe models \cite{Qiu:2011cy,Easson:2011zy,Cai:2012va,Osipov:2013ssa,Koehn:2013upa,Battarra:2014tga,Qiu:2015nha,Wan:2015hya,Banerjee:2016hom}, and also an attempt to describe the creation of a universe in the laboratory \cite{Rubakov:2013kaa}.

While, in the cosmological context, the energy density of Galileons can indeed increase in time in a healthy way, constructing a complete bouncing or Genesis cosmology (full evolution from $t=-\infty$ to $t=+\infty$) is a challenge. For example, one can construct a spatially flat bouncing model without pathologies at or near the bounce \cite{Easson:2011zy,Ijjas:2016tpn}, yet, in known examples, the gradient instabilities occur at some later or earlier epoch \cite{Cai:2012va,Koehn:2013upa,Battarra:2014tga,Qiu:2015nha}. Although these instabilities have been argued to remain under control due to higher derivative terms \cite{Koehn:2015vvy}, it would be interesting to design an example of a complete ``classical'' bouncing cosmological model without ghosts and gradient instabilities.

Another potential application of the NEC violation is a putative construction of stable asymptotically flat Lorentzian wormholes. However, previous attempts to design a wormhole supported by Galileon have failed \cite{Rubakov:2015gza,Rubakov:2016zah}.

It is worth noting that there exist bouncing models with nonzero spatial curvature \cite{Starobinsky:1980te,Ayon-Beato:2015eca}. Likewise, there are Lorentzian wormholes which are not asymptotically flat \cite{Ayon-Beato:2015eca}. These solutions employ scalar fields with fairly conventional kinetic terms that do not violate the NEC. On the contrary, we are interested in spatially flat bouncing cosmologies and asymptotically flat wormholes, which necessarily require NEC violation (hence our interest in Galileons).

Recently, two no-go theorems have been proven in the Galileon context \cite{Libanov:2016kfc,Rubakov:2016zah,Kobayashi:2016xpl}. Both apply to general relativity with the Galileon field and no other matter. One theorem shows that spatially flat bouncing cosmological solutions are always plagued with gradient instabilities \cite{Libanov:2016kfc,Kobayashi:2016xpl}. The other states that, in cubic Galileon theory, static spherically symmetric Lorentzian wormholes are always plagued with ghosts \cite{Rubakov:2016zah}. 

One might hope that these problems can be overcome by adding extra non-Galileonic matter. This matter, if it satisfies the NEC, must interact with Galileon directly; otherwise, the above theorems remain valid \cite{Libanov:2016kfc,Rubakov:2016zah,Kobayashi:2016xpl}. The simplest option is to add a scalar field with first-derivative terms in the Lagrangian. This is precisely the system studied in this paper. Somewhat surprisingly, we show that, at least for cubic Galileon, the above theorems are still at work: there are always gradient instabilities about bouncing cosmological solutions, and there always exist ghosts in backgrounds of static spherically symmetric Lorentzian wormholes.

Concerning wormholes, our spherically symmetric setting is not general. That is, we do not consider a cross term in a metric characteristic of Newman-Unti-Tamburino (NUT) spacetimes. In view of recent interesting results on NUT wormholes \cite{Ayon-Beato:2015eca,Clement:2015aka}, this generalization is worth pursuing in the future.

The paper is organized as follows. In Sec. \ref{sec:theory} we present the Lagrangian of the theory and obtain the expression for the stress-energy tensor. We also derive the equations of motion for Galileon and additional scalar field and quadratic Lagrangian for their perturbations. In Sec. \ref{sec:no-go_bounce} we consider spatially flat bouncing Friedmann-Lema\^{\i}tre-Robertson-Walker backgrounds and show that perturbations about them always have gradient instabilities at some epoch. In Sec. \ref{sec:no-go_wormhole} we study static, spherically symmetric Lorentzian wormholes and show that they are always plagued with ghosts. We conclude in Sec \ref{sec:concl}.

\section{The class of theories}
\label{sec:theory}
We study the class of theories with the Galileon field ${\pi}$ interacting with an additional scalar field $\phi$ and gravity. The Lagrangian of this theory in $(d + 2)$-dimensional space-time has the form
\begin{equation}
\label{eq:lagrangian}
\mathcal{L} = -\dfrac{1}{2\kappa}R + F(\pi,X,\phi,\tilde{X},Y) + K(\pi,X,\phi)\square\pi,
\end{equation}
where
\begin{equation*}
\begin{split}
X = (\partial\pi)^2 = g^{\mu\nu}\partial_{\mu}\pi\partial_{\nu}\pi,\quad \tilde{X} &= (\partial\phi)^2 = g^{\mu\nu}\partial_{\mu}\phi\partial_{\nu}\phi,\quad Y = g^{\mu\nu}\partial_{\mu}\pi\partial_{\nu}\phi, \\
\square\pi &= g^{\mu\nu}\nabla_{\mu}\nabla_{\nu}\pi,
\end{split}
\end{equation*}
$R$ is the scalar curvature, and $\kappa = 8\pi G$. We do not study the case in which the function $K$ depends on $\tilde{X}$ and/or $Y$ since the equation of motion for $\pi$ would contain the third derivatives of $\phi$.

The stress-energy tensor\footnote{We use the mostly negative metric $\{+-,\cdots,-\}$.} corresponding to the Lagrangian (\ref{eq:lagrangian}) reads
\begin{equation}
\label{eq:stress_energ_tens}
\begin{split}
T_{\mu\nu} =
&-g_{\mu\nu}F + 2F_X \partial_\mu\pi\partial_\nu\pi + 2F_{\tilde{X}} \partial_\mu\phi\partial_\nu\phi + 2F_Y\partial_\mu\pi\partial_\nu\phi \\
&+g_{\mu\nu}K_\pi \partial_\rho\pi\partial^\rho\pi + g_{\mu\nu}K_\phi\partial_\rho\pi\partial^\rho\phi - 2K_\pi\partial_\mu\pi\partial_\nu\pi - 2K_\phi\partial_\mu\pi\partial_\nu\phi \\
&+ 2g_{\mu\nu}K_X\nabla^\rho\nabla^\lambda\pi\partial_\rho\pi\partial_\lambda\pi + 2K_X\square\pi\partial_\mu\pi\partial_\nu\pi - 4K_X\nabla_\mu\nabla^\rho\pi\partial_\rho\pi\partial_\nu\pi.
\end{split}
\end{equation}
One obtains one equation of motion by varying the field $\pi$ in the Lagrangian (\ref{eq:lagrangian}),
\begin{equation}
\begin{split}
\label{eq:pi_feild_eq}
&\nabla_\mu\nabla^\nu\pi[-4F_{XX}\partial^\mu\pi\partial_\nu\pi - 4F_{XY}\partial^\mu\pi\partial_\nu\phi - F_{YY}\partial^\mu\phi\partial_\nu\phi \\
&+4K_{\pi X}\partial^\mu\pi\partial_\nu\pi + 2K_X\nabla^\mu\nabla_\nu\pi + 4K_{XX}\nabla^\mu\nabla^\rho\pi\partial_\rho\pi\partial_\nu\pi + 4K_{X \phi}\partial^\mu\pi\partial_\nu\phi] \\+
&\nabla_\mu\nabla^\mu\pi[-2F_X + 2K_\pi - 2K_{\pi X}\partial_\mu\pi\partial^\mu\pi - 2K_X\square\pi - 2K_{X \phi}\partial_\mu\pi\partial^\mu\phi] \\-
&\nabla_\mu\nabla^\nu\pi \cdot \nabla_\rho\nabla^\rho\pi \cdot 4K_{XX}\partial^\mu\pi\partial_\nu\pi \\+
&\nabla_\mu\nabla^\nu\phi[-4F_{X\tilde{X}}\partial^\mu\pi\partial_\nu\phi - 2F_{XY}\partial^\mu\pi\partial_\nu\pi - 2F_{\tilde{X}Y}\partial^\mu\phi\partial_\nu\phi - F_{YY}\partial^\mu\pi\partial_\nu\phi] \\+
&\nabla_\mu\nabla^\mu\phi[-F_Y + K_\phi] + 2K_X R_{\mu\nu}\partial^\mu\pi\partial^\nu\pi + \ldots = 0.
\end{split}
\end{equation}
Hereafter, dots denote terms without second derivatives, and
\begin{equation*}
F_\pi = \dfrac{\partial F}{\partial \pi},\quad F_X = \dfrac{\partial F}{\partial X},\quad F_\phi = \dfrac{\partial F}{\partial \phi},\quad F_{\tilde{X}} = \dfrac{\partial F}{\partial \tilde{X}}, \quad F_Y = \dfrac{\partial F}{\partial Y}
\end{equation*} 
and similarly for the function $K$. The equation of motion obtained by varying the field $\phi$ has the form
\begin{equation}
\begin{split}
\label{eq:phi_feild_eq}
&\nabla_\mu\nabla^\nu\pi[-4F_{X\tilde{X}}\partial^\mu\pi\partial_\nu\phi - 2F_{XY}\partial^\mu\pi\partial_\nu\pi - 2F_{\tilde{X}Y}\partial^\mu\phi\partial_\nu\phi - F_{YY}\partial^\mu\pi\partial_\nu\phi] \\+
&\nabla_\mu\nabla^\mu\pi[-F_Y + K_\phi] \\+
&\nabla_\mu\nabla^\nu\phi[-4F_{\tilde{X}\tilde{X}}\partial^\mu\phi\partial_\nu\phi - 4F_{\tilde{X}Y}\partial^\mu\pi\partial_\nu\phi - F_{YY}\partial^\mu\pi\partial_\nu\pi] \\-
&\nabla_\mu\nabla^\mu\phi\cdot 2F_{\tilde{X}} + \ldots = 0.
\end{split}
\end{equation}

Let us consider the small perturbations $\chi = \chi\left(x^\mu\right)$ and $\varphi = \varphi\left(x^\mu\right)$ about solutions to the field equations $\pi_c$ and $\phi_c$, respectively. By substituting $\pi = \pi_c + \chi$ and $\phi = \phi_c + \varphi$ into Eq. (\ref{eq:pi_feild_eq}) and omitting the subscript $c$, one obtains the equation of motion for the perturbations $\chi$ and $\varphi$. We are interested in high momentum and frequency modes; therefore, we retain only those terms with second derivatives and write 
\begin{equation}
\label{eq:chi_feild_eq}
\begin{split}
&\nabla_\mu\nabla^\nu\chi[-4F_{XX}\partial^\mu\pi\partial_\nu\pi - 4F_{XY}\partial^\mu\pi\partial_\nu\phi - F_{YY}\partial^\mu\phi\partial_\nu\phi \\
&+ 4K_{\pi X}\partial^\mu\pi\partial_\nu\pi + 4K_X\nabla^\mu\nabla_\nu\pi - 4K_{XX}\square\pi\partial^\mu\pi\partial_\nu\pi \\
&+ 8K_{XX}\nabla^\mu\nabla^\rho\pi\partial_\rho\pi\partial_\nu\pi + 4K_{X \phi}\partial^\mu\pi\partial_\nu\phi] \\+
&\nabla_\mu\nabla^\mu\chi[-2F_X + 2K_\pi - 2K_{\pi X}(\partial\pi)^2 \\
&- 4K_X\square\pi - 4K_{XX}\nabla_\mu\nabla^\nu\pi\partial^\mu\pi\partial_\nu\pi - 2K_{X\phi}\partial_\mu\pi\partial^\mu\phi] \\+
&\nabla_\mu\nabla^\nu\varphi[-4F_{X\tilde{X}}\partial^\mu\pi\partial_\nu\phi - 2F_{XY}\partial^\mu\pi\partial_\nu\pi - 2F_{\tilde{X}Y}\partial^\mu\phi\partial_\nu\phi - F_{YY}\partial^\mu\pi\partial_\nu\phi] \\+
&\nabla_\mu\nabla^\mu\varphi[-F_Y + K_\phi]
+ 2K_XR^{(1)}_{\mu\nu}\partial^\mu\pi\partial^\nu\pi + \ldots = 0.
\end{split}
\end{equation}
Here, $R^{(1)}_{\mu\nu}$ involves second derivatives of the metric perturbations. Following Ref. \cite{Deffayet:2010qz}, we integrate the metric perturbations out by making use of the Einstein equations $R_{\mu\nu} - \frac{1}{2}g_{\mu\nu}R = \kappa T_{\mu\nu}$. By substituting $\pi = \pi_c + \chi$ into the expression for the stress-energy tensor (\ref{eq:stress_energ_tens}), one obtains the following relation drawn from the linearized Einstein equations:
\begin{equation*}
2K_XR^{(1)}_{\mu\nu}\partial^\mu\pi\partial^\nu\pi = 2K_X^2 \kappa \Big[\dfrac{2(d - 1)}{d} (\partial\pi)^4 \square\chi - 4(\partial\pi)^2\partial^\mu\pi\partial_\nu\pi\nabla_\mu\nabla^\nu\chi\Big] + \ldots,
\end{equation*}
which is used in Eq. (\ref{eq:chi_feild_eq}) to get rid of the second derivatives of the metric perturbations. Linearized equation (\ref{eq:phi_feild_eq}) reads
\begin{equation}
\begin{split}
\label{eq:varphi_feild_eq}
&\nabla_\mu\nabla^\nu\chi[-4F_{X\tilde{X}}\partial^\mu\pi\partial_\nu\phi - 2F_{XY}\partial^\mu\pi\partial_\nu\pi - 2F_{\tilde{X}Y}\partial^\mu\phi\partial_\nu\phi - F_{YY}\partial^\mu\pi\partial_\nu\phi] \\+
&\nabla_\mu\nabla^\mu\chi[-F_Y + K_\phi] \\+
&\nabla_\mu\nabla^\nu\varphi[-4F_{\tilde{X}\tilde{X}}\partial^\mu\phi\partial_\nu\phi -4F_{\tilde{X}Y}\partial^\mu\pi\partial_\nu\phi - F_{YY}\partial^\mu\pi\partial_\nu\pi] \\-
&\nabla_\mu\nabla^\mu\varphi \cdot 2F_{\tilde{X}} + \ldots = 0.
\end{split}
\end{equation}
Now we exploit the equations of motion (\ref{eq:chi_feild_eq}) and (\ref{eq:varphi_feild_eq}) to construct the quadratic Lagrangian for perturbations:
\begin{equation}
\begin{split}
\label{eq:pertrub_lagrangian}
\mathcal{L}^{(2)}_{\chi\varphi} =
&\partial_\mu\chi\partial^\nu\chi\Big[2F_{XX}\partial^\mu\pi\partial_\nu\pi + 2F_{XY}\partial^\mu\pi\partial_\nu\phi + \frac{1}{2}F_{YY}\partial^\mu\phi\partial_\nu\phi \\
&- 2K_{\pi X}\partial^\mu\pi\partial_\nu\pi - 2K_X\nabla^\mu\nabla_\nu\pi + 2K_{XX}\square\pi\partial^\mu\pi\partial_\nu\pi \\
&-4K_{XX}\nabla^\mu\nabla^\rho\pi\partial_\rho\pi\partial_\nu\pi - 2K_{X\phi}\partial^\mu\pi\partial_\nu\phi\Big] \\+
&\partial_\mu\chi\partial^\mu\chi\Big[F_X - K_\pi + K_{\pi X}(\partial\pi)^2 + 2K_X\square\pi \\
&+ 2K_{XX}\nabla_\mu\nabla^\nu\pi\partial^\mu\pi\partial_\nu\pi + K_{X\phi}\partial_\mu\pi\partial^\mu\phi\Big] \\+
&\partial_\mu\chi\partial^\nu\varphi\Big[4F_{X\tilde{X}}\partial^\mu\pi\partial_\nu\phi + 2F_{XY}\partial^\mu\pi\partial_\nu\pi + 2F_{\tilde{X}Y}\partial^\mu\phi\partial_\nu\phi + F_{YY}\partial^\mu\pi\partial_\nu\phi\Big] \\+
&\partial_\mu\chi\partial^\mu\varphi\Big[F_Y - K_\phi\Big] \\+
&\partial_\mu\varphi\partial^\nu\varphi\Big[2F_{\tilde{X}\tilde{X}}\partial^\mu\phi\partial_\nu\phi + 2F_{\tilde{X}Y}\partial^\mu\pi\partial_\nu\phi + \frac{1}{2}F_{YY}\partial^\mu\pi\partial_\nu\pi\Big] \\+
&\partial_\mu\varphi\partial^\mu\varphi \cdot F_{\tilde{X}} - K^2_X\kappa\left[\dfrac{2(d - 1)}{d}(\partial\pi)^4(\partial\chi)^2 - 4(\partial\pi)^2\partial^\mu\pi\partial_\nu\pi\partial_\mu\chi\partial^\nu\chi\right].
\end{split}
\end{equation}
Let us now use this Lagrangian to study the stability of cosmological bounces and Lorentzian wormholes.

\section{No-go for cosmological bounce}
\label{sec:no-go_bounce}
In this section we prove the existence of gradient instabilities in spatially flat bouncing cosmological models in the theory (\ref{eq:lagrangian}). The background fields $\pi(t)$ and $\phi(t)$ are spatially homogeneous and the metric is
\begin{equation*}
\mathrm{d}s^2 = \mathrm{d}t^2 - a^2(t)\gamma_{ij}\mathrm{d}x^i\mathrm{d}x^j,
\end{equation*}
where $\gamma_{ij} = \delta_{ij}$ is the metric of $(d + 1)$-dimensional flat space.

Equation (\ref{eq:stress_energ_tens}) leads to expressions for the energy density and the pressure of the Galileon and an additional scalar field,
\begin{subequations}
\label{eq:rho&p_bounce}
\begin{align}
&T^0_0 = \rho = -F + 2F_X\dot{\pi}^2 + 2F_{\tilde{X}}\dot{\phi}^2 + 2F_Y \dot{\pi}\dot{\phi} - K_{\pi}\dot{\pi}^2 - K_{\phi}\dot{\pi}\dot{\phi} + 2(d+1)K_X H \dot{\pi}^3, \\-
&T^i_j = p\cdot\delta^i_j = (F -  K_{\pi}\dot{\pi}^2 - K_\phi\dot{\pi}\dot{\phi} - 2K_X \dot{\pi}^2\ddot{\pi})\cdot\delta^i_j,
\end{align}
\end{subequations}
where $H = \dfrac{\dot{a}}{a}$ is the Hubble parameter. The quadratic Lagrangian (\ref{eq:pertrub_lagrangian}) reads
\begin{equation}
\label{eq:pertrub_lagrangian_lorenzsum_bounce}
\mathcal{L}^{(2)}_{\chi\varphi} = A_\pi\dot{\chi}^2 - B_\pi\left(\dfrac{\partial_i\chi}{a}\right)^2 + A_\phi\dot{\varphi}^2 - B_\phi\left(\dfrac{\partial_i\varphi}{a}\right)^2 + 2C_{A}\dot{\chi}\dot{\varphi} - 2C_{B}\dfrac{\partial_i\chi\partial_i\varphi}{a^2},
\end{equation}
where
\begin{subequations}
\begin{align}
\label{eq:B_pi}
B_\pi
&= F_X - K_\pi + K_{\pi X}\dot{\pi}^2 + 2K_X\ddot{\pi} + 2K_{XX}\dot{\pi}^2\ddot{\pi} + K_{X\phi}\dot{\pi}\dot{\phi}\\ \nonumber 
&+ 2dK_X H\dot{\pi} - \dfrac{2(d - 1)}{d}\kappa K^2_X\dot{\pi}^4, \\
\label{eq:B_phi}
B_\phi
&= F_{\tilde{X}}, \\
\label{eq:C_B}
C_B
&= \frac{1}{2}(F_Y - K_\phi),
\end{align}
\end{subequations}
and we do not need the expressions for $A_\pi$, $A_\phi$, and $C_A$ in what follows.
The Friedmann equation and the covariant stress-energy conservation give
\begin{equation*}
\begin{aligned}
&H^2 = \dfrac{2}{d(d + 1)} \kappa\rho, \\
&\dot{\rho} = -(d + 1)H(\rho + p), 
\end{aligned}
\end{equation*}
so that
\begin{equation}
\label{eq:NEC_combination_bounce}
\rho + p = -\dfrac{d}{\kappa}\dot{H}.
\end{equation}
 It follows from Eqs. (\ref{eq:rho&p_bounce}), (\ref{eq:B_pi}), and (\ref{eq:NEC_combination_bounce}) that
\begin{equation}
\label{eq:B(Q)}
2B_\pi\dot{\pi}^2 = \dot{Q} - \dfrac{2(d - 1)}{d}\kappa K_X\dot{\pi}^3Q + 2\left[\left(K_\phi - F_Y\right)\dot{\pi}\dot{\phi} - F_{\tilde{X}}\dot{\phi}^2\right],
\end{equation}
where 
\begin{equation*}
Q = 2K_X\dot{\pi}^3 - \dfrac{d}{\kappa}H.
\end{equation*}
Note that the latter combination has been used in Ref. \cite{Libanov:2016kfc}.
It is useful to write the Lagrangian for perturbations (\ref{eq:pertrub_lagrangian_lorenzsum_bounce}) in the matrix form:
\begin{equation*}
\mathcal{L}^{(2)}_{\chi\varphi} = \dot{\psi}^{T}M_A \dot{\psi} - \dfrac{\partial_i\psi^{T}}{a}M_B\dfrac{\partial_i\psi}{a},
\end{equation*}
where
\begin{equation*}
\psi =
\begin{pmatrix}
\chi \\
\varphi
\end{pmatrix}, \qquad
M_A =
\begin{pmatrix}
A_\pi & C_A    \\
C_A   & A_\phi\\
\end{pmatrix}, \qquad
M_B =
\begin{pmatrix}
B_\pi & C_B    \\
C_B   & B_\phi\\
\end{pmatrix}.
\end{equation*}
Gradient instabilities are absent if and only if the matrix $M_B$ is positive definite,
\begin{subequations}
\begin{align}
&B_\pi > 0, \\
\label{eq:B_phi>0}
&B_\phi = F_{\tilde{X}} > 0, \\
&\mathrm{det}(M_B) = B_\pi \cdot B_\phi - C_B^2 > 0.
\end{align}
\end{subequations}
According to Eqs. (\ref{eq:B_pi}), (\ref{eq:B_phi}), (\ref{eq:C_B}), and (\ref{eq:B(Q)}), the condition $\mathrm{det}(M_B)>0$ reads
\begin{equation*}
\frac{1}{2\dot{\pi}^2} \left[ \dot{Q} - \dfrac{2(d - 1)}{d}\kappa K_x\dot{\pi}^3Q\right]\cdot F_{\tilde{X}} + (K_\phi - F_Y)\cdot F_{\tilde{X}}\frac{\dot{\phi}}{\dot{\pi}} - \left(F_{\tilde{X}}\right)^2\left(\frac{\dot{\phi}}{\dot{\pi}}\right)^2 - \frac{1}{4}\left(K_\phi - F_Y\right)^2 >0,
\end{equation*}
or, equivalently,
\begin{equation}
\label{eq:grad_stability_cond}
2F_{\tilde{X}}\cdot\left[\dot{Q} - \dfrac{2(d - 1)}{d}\kappa K_x\dot{\pi}^3Q\right] > \left[2F_{\tilde{X}}\dot{\phi} - (K_\phi - F_Y)\dot{\pi}\right]^2.
\end{equation}
The inequality (\ref{eq:grad_stability_cond}), together with (\ref{eq:B_phi>0}), requires
\begin{equation}
\label{eq:Q_bounce}
\dot{Q} - \dfrac{2(d - 1)}{d}\kappa K_x\dot{\pi}^3Q > 0.
\end{equation}
This is the same inequality as in the theory with a single Galileon \cite{Libanov:2016kfc}. It cannot be satisfied at all times in the bouncing model. Indeed, let us introduce the combination $R = Q/a^{d - 1}$. In terms of this combination, the inequality (\ref{eq:Q_bounce}) reads
\begin{equation*}
\dot{R} - \dfrac{d - 1}{d}\kappa a^{d - 1}R^2 \geq 0.
\end{equation*}
Upon integration from $t_i$ to $t_f \geq t_i$, one obtains the inequality
\begin{equation}
\label{eq:R}
\dfrac{1}{R(t_i)} - \dfrac{1}{R(t_f)} \geq \dfrac{d - 1}{d} \kappa \int_{t_i}^{t_f}dt \, a^{d - 1}. 
\end{equation}
Suppose that $R(t_i) \geq 0$. Taking into account that, in bouncing cosmology, $t$ runs from $-\infty$ to $+\infty$ and $a(t)$ is bounded from below, one finds that $R^{-1}(t)$ necessarily crosses zero, which signals the presence of a singular point $R(t_*) = \infty$ during the evolution. The possibility that $R(t)$ is negative at all times is also inconsistent with (\ref{eq:R}) for nonsingular $R(t)$'s.

\section{No-go for wormhole}
\label{sec:no-go_wormhole}
A similar argument works for wormholes. The basic difference with the
bounce is the permutation of $t$ and $r$; hence, as we now show, there are ghosts about static, spherically symmetric wormholes rather than gradient instabilities.

A wormhole with the fields $\pi$ and $\phi$ is described by the static and spherically symmetric solution $\pi(r)$ and $\phi(r)$ to the field equations (\ref{eq:pi_feild_eq}) and (\ref{eq:phi_feild_eq}), and the following metric:
\begin{equation*}
\mathrm{d}s^2 = a^2(r)\mathrm{d}t^2 - \mathrm{d}r^2 - c^2(r)\gamma_{\alpha\beta}\mathrm{d}x^\alpha \mathrm{d}x^\beta,
\end{equation*}
where $x^\alpha$ and $\gamma_{\alpha\beta}$ are the cordinates and the metric on a unit $d$-dimensional sphere. The coordinate $r$ runs from $-\infty$ to $+\infty$, and the metric coefficients are strictly positive and bounded from below:
\begin{equation}
\label{eq:a&c}
a(r) \geq a_{min} > 0, \quad c(r) \geq R_{min} > 0,
\end{equation}
where $R_{min}$ is the radius of the wormhole throat. \\

The components of the stress-energy tensor (\ref{eq:stress_energ_tens}) are
\begin{subequations}
\label{eq:rho&p_wormhole}
\begin{align}
\label{eq:rho}
T^0_0
&= \rho = -F -  K_{\pi}(\pi')^2 - K_\phi\pi'\phi' + 2K_X (\pi')^2\pi'', \\\label{eq:p} -T^r_r
&= p_r = F + 2F_X(\pi')^2 + 2F_{\tilde{X}}(\phi')^2 + 2F_Y \pi'\phi' \nonumber \\
&- K_{\pi}(\pi')^2 - K_{\phi}\pi'\phi' - 2K_X (\pi')^3\left[\dfrac{a'}{a} + d\dfrac{c'}{c} \right], \\ -T^\alpha_\beta
& = p_t \cdot \delta^\alpha_\beta = -T^0_0 \cdot \delta^\alpha_\beta,
\end{align}
\end{subequations}
where prime denotes $\mathrm{d}/\mathrm{d}r$. Now we write the Lagrangian for the perturbations,
\begin{equation}
\begin{split}
\label{eq:pertrub_lagrangian_lorenzsum_wormhole}
\mathcal{L}^{(2)}_{\chi\varphi} =
&\mathcal{A}_\pi\left(\dfrac{\dot{\chi}}{a}\right)^2 - \mathcal{B}_\pi\left(\chi'\right)^2 -\mathcal{D}_\pi\gamma^{\alpha\beta}\dfrac{\partial_\alpha\chi\partial_\beta\chi}{c^2} \\+
&\mathcal{A}_\phi\left(\dfrac{\dot{\varphi}}{a}\right)^2 - \mathcal{B}_\phi\left(\varphi'\right)^2 -\mathcal{D}_\phi\gamma^{\alpha\beta}\dfrac{\partial_\alpha\varphi\partial_\beta\varphi}{c^2} \\+
&2\mathcal{C}_{\mathcal{A}}\dfrac{\dot{\chi}\dot{\varphi}}{a^2} - 2\mathcal{C}_{\mathcal{B}}\chi'\varphi' - 2C_\mathcal{D}\gamma^{\alpha\beta}\dfrac{\partial_\alpha\chi\partial_\beta\varphi}{c^2},
\end{split}
\end{equation}
where
\begin{subequations}
\begin{align}
\label{eq:A_pi}
\mathcal{A}_\pi
&= F_X - K_\pi - K_{\pi X}(\pi')^2 - 2K_X\pi'' + 2K_{XX}(\pi')^2\pi'' - K_{X \phi}\pi'\phi' \\ \nonumber 
&- 2dK_X \dfrac{c'}{c}\pi' - \dfrac{2(d - 1)}{d}\kappa K^2_X(\pi')^4, \\
\label{eq:A_phi}
\mathcal{A}_\phi
&=F_{\tilde{X}},\\
\label{eq:C_A}
\mathcal{C}_\mathcal{A}
&= \frac{1}{2}(F_Y - K_\phi),
\end{align}
\end{subequations}
and we do not need the expressions for $\mathcal{B}_\pi$, $\mathcal{B}_\phi$, $\mathcal{C}_\mathcal{B}$ and $\mathcal{D}_\pi$, $\mathcal{D}_\phi$, $\mathcal{C}_\mathcal{D}$. A combination of the Einstein equations gives \cite{Rubakov:2016zah}
\begin{equation}
\label{eq:NEC_combination_wormhole}
T^0_0 - T^r_r = \rho + p_r = -\dfrac{d}{\kappa}\dfrac{a}{c}\left(\frac{c'}{a}\right)'.
\end{equation}
It follows from Eqs. (\ref{eq:rho}), (\ref{eq:p}), (\ref{eq:A_pi}), and (\ref{eq:NEC_combination_wormhole}) that
\begin{equation}
\label{eq:A(Q)}
2\mathcal{A}_\pi(\pi')^2\dfrac{c}{a} = -\mathcal{Q}' - \dfrac{2(d - 1)}{d}\kappa K_X (\pi')^3\mathcal{Q} + 2\Big[\left(K_\phi - F_Y\right)\pi'\phi' - F_{\tilde{X}}(\phi')^2\Big]\dfrac{c}{a},
\end{equation}
where
\begin{equation*}
\mathcal{Q} = 2\dfrac{c}{a}K_X(\pi')^3 + \dfrac{d}{\kappa}\dfrac{c'}{a}.
\end{equation*}
Note that the latter combination was used in Ref. \cite{Rubakov:2016zah}. We again rewrite the Lagrangian for perturbations (\ref{eq:pertrub_lagrangian_lorenzsum_wormhole}) in the matrix form:
\begin{equation*}
\mathcal{L}^{(2)}_{\chi\varphi} = \dfrac{\dot{\psi}^{T}}{a}M_\mathcal{A}\dfrac{\dot{\psi}}{a} - (\psi')^{T}M_{\mathcal{B}}\psi' - \gamma^{\alpha\beta}\dfrac{\partial_\alpha\psi^T}{c}M_\mathcal{D}\dfrac{\partial_\beta\psi}{c},
\end{equation*}
where
\begin{equation*}
\psi =
\begin{pmatrix}
\chi \\
\varphi
\end{pmatrix}, \qquad
M_\mathcal{A}=
\begin{pmatrix}
\mathcal{A}_\pi & C_\mathcal{A}    \\
C_\mathcal{A}   & \mathcal{A}_\phi\\
\end{pmatrix}, \qquad
M_\mathcal{B}=
\begin{pmatrix}
\mathcal{B}_\pi & C_\mathcal{B}    \\
C_\mathcal{B}   & \mathcal{B}_\phi\\
\end{pmatrix}, \qquad
M_\mathcal{D}=
\begin{pmatrix}
\mathcal{D}_\pi & C_\mathcal{D}    \\
C_\mathcal{D}   & \mathcal{D}_\phi\\
\end{pmatrix}.
\end{equation*}
Ghost instabilities are absent if and only if the matrix $M_\mathcal{A}$ is positive definite. Therefore, for an absence of ghosts, we have to impose
\begin{subequations}
\begin{align}
&\mathcal{A}_\pi > 0, \\
\label{eq:A_phi>0}
&\mathcal{A}_\phi = F_{\tilde{X}} > 0, \\
&\mathrm{det}(M_\mathcal{A}) = \mathcal{A}_\pi \cdot \mathcal{A}_\phi - C_\mathcal{A}^2 > 0.
\end{align}
\end{subequations}
According to Eqs. (\ref{eq:A_pi}), (\ref{eq:A_phi}), (\ref{eq:C_A}), and (\ref{eq:A(Q)}), the condition $\mathrm{det}(M_\mathcal{A})>0$ reads
\begin{equation*}
\begin{split}
\frac{1}{2(\pi')^2\dfrac{c}{a}} \left[ -\mathcal{Q}' - \dfrac{2(d - 1)}{d}\kappa K_X(\pi')^3\mathcal{Q}\right]\cdot F_{\tilde{X}}
&+ (K_\phi - F_Y)\cdot F_{\tilde{X}}\frac{\phi'}{\pi'} \\
&- \left(F_{\tilde{X}}\right)^2\left(\frac{\phi'}{\pi'}\right)^2 - \frac{1}{4}\left(K_\phi - F_Y\right)^2 >0,
\end{split}
\end{equation*}
i. e.,
\begin{equation}
\label{eq:ghost_stability_cond}
2F_{\tilde{X}}\cdot\dfrac{a}{c}\cdot\left[-\mathcal{Q}' - \dfrac{2(d - 1)}{d}\kappa K_X(\pi')^3\mathcal{Q}\right] > \Big[2F_{\tilde{X}}\phi'-(K_\phi - F_Y)\pi'\Big]^2.
\end{equation}
The inequality (\ref{eq:ghost_stability_cond}), together with (\ref{eq:a&c}) and (\ref{eq:A_phi>0}), requires
\begin{equation}
\label{eq:Q_wormhole}
-\mathcal{Q}' - \dfrac{2(d - 1)}{d}\kappa K_X(\pi')^3\mathcal{Q} > 0.
\end{equation}
This is the same inequality as in the theory with a single Galileon \cite{Rubakov:2016zah}. It cannot be satisfied for all $r$'s. Indeed, in terms of the combination $\mathcal{R} = \mathcal{Q}/c^{d - 1}$, the inequality (\ref{eq:Q_wormhole}) reads
\begin{equation*}
-\mathcal{R}' - \dfrac{d - 1}{d} \kappa a c^{d - 2}\mathcal{R}^2 \geq 0.
\end{equation*}
This inequality cannot be satisfied for nonsingular $\mathcal{R}$ for the same reason as in Sec. \ref{sec:no-go_bounce}.

\section{Conclusion}
\label{sec:concl}
In this paper we generalized the no-go arguments \cite{Libanov:2016kfc,Rubakov:2016zah,Kobayashi:2016xpl} for bounces and wormholes to the case of a Galileon interacting with another scalar field. We saw that adding of extra scalar field modifies the relevant inequalities and makes them even stronger. The proofs of the no-go theorems are of a technical character and the physical reason behind them is unclear. Note, however, that the absence of stable wormholes can be interpreted as censorship against time machines, while bounce is a cosmological analog of a wormhole. In this sense, both no-go theorems have an intrinsic relationship with the inability to build a time machine.
 
\section*{Acknowledgments}
We are indebted to V. Rubakov for the useful discussions and the thoughtful reading of the manuscript. The authors are grateful to M. Libanov and A. Sosnovikov for their valuable comments. This work has been supported by Russian Science Foundation Grant No. 14-22-00161.

\end{document}